\documentclass[runningheads]{llncs}

\usepackage{eccv}
\usepackage{tabularx}
\usepackage{adjustbox}

\usepackage{eccvabbrv}

\usepackage{graphicx}
\usepackage{booktabs}

\usepackage[accsupp]{axessibility}  

\usepackage[linesnumbered,ruled,vlined]{algorithm2e}
\usepackage{amsmath, amssymb}
\usepackage{caption}
\usepackage{subcaption}
\usepackage{multicol}

\usepackage[pagebackref,breaklinks,colorlinks]{hyperref}

\usepackage{orcidlink}

\begin{document}

\title{Solving the inverse problem of microscopy deconvolution with a residual Beylkin-Coifman-Rokhlin neural network} 

\titlerunning{m-rBCR microscopy}

\author{Rui Li\inst{1,2}\orcidlink{0000-1111-2222-3333} \and
Mikhail Kudryashev\inst{3,4}\orcidlink{1111-2222-3333-4444} \and
Artur Yakimovich\inst{1,2,5}\orcidlink{2222--3333-4444-5555}}

\authorrunning{Li et al.}

\institute{Center for Advanced Systems Understanding (CASUS), Görlitz, Germany \\ 
\and
Helmholtz-Zentrum Dresden-Rossendorf e. V. (HZDR), Dresden, Germany \\ \and
Max Delbrück Center for Molecular Medicine in the Helmholtz Association, Berlin, Germany \\ 
\and
Institute of Medical Physics and Biophysics, Charite-Universitätsmedizin, Berlin, Germany \\ 
\and
Institute of Computer Science, University of Wrocław, Wrocław, Poland
\email{a.yakimovich@hzdr.de}}

\maketitle

\begin{abstract}
  Optic deconvolution in light microscopy (LM) refers to recovering the object details from images, revealing the ground truth of samples. Traditional explicit methods in LM rely on the point spread function (PSF) during image acquisition. Yet, these approaches often fall short due to inaccurate PSF models and noise artifacts, hampering the overall restoration quality. In this paper, we approached the optic deconvolution as an inverse problem. Motivated by the nonstandard-form compression scheme introduced by Beylkin, Coifman, and Rokhlin (BCR), we proposed an innovative physics-informed neural network Multi-Stage Residual-BCR Net (m-rBCR) to approximate the optic deconvolution. We validated the m-rBCR model on four microscopy datasets - two simulated microscopy datasets from ImageNet and BioSR, real dSTORM microscopy images, and real widefield microscopy images. In contrast to the explicit deconvolution methods (e.g. Richardson-Lucy) and other state-of-the-art NN models (U-Net, DDPM, CARE, DnCNN, ESRGAN, RCAN, Noise2Noise, MPRNet, and MIMO-U-Net), the m-rBCR model demonstrates superior performance to other candidates by PSNR and SSIM  in two real microscopy datasets and the simulated BioSR dataset. In the simulated ImageNet dataset, m-rBCR ranks the second-best place (right after MIMO-U-Net). With the backbone from the optical physics, m-rBCR exploits the trainable parameters with better performances (from $\sim$30 times fewer than the benchmark MIMO-U-Net to $\sim$210 times than ESRGAN). This enables m-rBCR to achieve a shorter runtime (from $\sim$3 times faster than MIMO-U-Net to $\sim$300 times faster than DDPM). To summarize, by leveraging physics constraints our model reduced potentially redundant parameters significantly in expertise-oriented NN candidates and achieved high efficiency with superior performance.
\end{abstract}

\section{Introduction}

Light microscopy plays a vital role in examining biological specimens. The optical lens system captures sample information as images through the operation optic convolution \cite{mcnally_three-dimensional_1999, sibarita_deconvolution_2005}. This process is mathematically expressed as $I(x)=A(t)\ast o(x)$. $o$ denotes the object information. $I$ represents the microscopy images and the operator $A$ stands for the convolution operator specific to the optical physics model. The convolution kernel, commonly referred to as the Point Spread Function ($PSF$) in LM \cite{aguet_super-resolution_2009}, plays a crucial role in image acquisition. A more specific representation of $A$ is denoted as $A(PSF(t))$. In reality, hardware limitations and operational variances introduce disruptive elements $\varepsilon$ during imaging. This revises the model into $I(x) = A(PSF(t))\ast o(x) + \varepsilon$. The ultimate goal is to extract the true sample information $o(x)$ from the microscopy image $I$. This process is formally defined as optic deconvolution. \cite{swedlow_live_2002}. 

Deconvolution is critically important to reveal detailed information in microscopy images. Several previous approaches have explored optic deconvolution from physics perspectives \cite{mcnally_three-dimensional_1999, sibarita_deconvolution_2005, kenig_blind_2010, noauthor_blind_nodate}. The deconvolution process hinges on the physical model of the $PSF$. Traditional deconvolution methods, like the Richardson-Lucy model, require an explicit formulation of the $PSF$. Yet, these methods face challenges arising from inaccurate physical models of the $PSF$ and disturbances during image acquisition, such as noise and artifacts. In many cases where $PSF$ information is unavailable, blind deconvolution becomes essential \cite{lim_blind_2019,  noauthor_blind_nodate}.

Currently, data-driven approaches like Deep learning (DL) show promise in many microscopy applications, such as target segmentation \cite{navab_u-net_2015}, super-resolution microscopy \cite{hatamizadeh_unetr_2021} and 3D microscopy \cite{li_weak-labelling_2023}. In the realm of resolution enhancement, DL models, such as CARE \cite{weigert_content-aware_2018}, DnCNN \cite{zhang_beyond_2017}, Noise2Noise \cite{lehtinen_noise2noise_2018}, MIMO-U-Net \cite{cho_rethinking_2021}, RCAN \cite{zhang_image_2018}, MPRNet \cite{zamir_multi-stage_2021}, U-Net \cite{navab_u-net_2015}, DDPM \cite{ho_denoising_2020} and ESRGAN \cite{wang_esrgan_2018}, allow recovering of high-resolution details from low-resolution inputs. Even though deblurring/denoising does not precisely replicate optical convolution, these modules achieve good performance in microscopy use cases. Yet, the majority of deep learning models for deconvolution neglect the physical texture of the problems. This leads to an increase in the model size in pursuit of improved performance. A few approaches have attempted to tackle data-driven blind deconvolution with a backbone of physical models. Tal and colleagues \cite{sibarita_deconvolution_2005} proposed a blind deconvolution method using traditional machine learning. They utilized specially designed Principal Component Analysis (PCA) to learn the PSF space. The model then sampled an appropriate PSF during deconvolution. Lim and colleagues \cite{lim_blind_2019} designed a convolution neural network (CNN) model embedded with the PSF information inside the model. 

In this work, we formulate deconvolution as an inverse problem \cite{bal_introduction_nodate, bertero_introduction_2021} for solving an integral operator. Based on the wavelet representation proposed by Beylkin, Coifman, and Rokhlin (BCR) \cite{beylkin_fast_1991, fan_bcr-net_2019}, we approximate the inverse process with a physics-informed Neural Network (PINN) model. Our proposed multi-stage residual BCR net (m-rBCR) is grounded in physical insights for deconvolution operation. Validated on four datasets (two simulated microscopy images datasets and two real microscopy datasets), the m-rBCR model demonstrates superior performance with significantly fewer trainable parameters (up to $\sim$210 times) and shorter run-time (up to $\sim$300 times) compared to the other state-of-the-art models (RL, U-Net, DDPM, MIMO-U-Net, Noise2Noise, CARE, DnCNN, ESRGAN, MPRNet and RCAN).

\section{Related work}

\subsection{Optical Models for Microscopy}
In a more specific form, Eq. \ref{convPhysic} \cite{sibarita_deconvolution_2005} denotes the previously mentioned optic convolution process. $O(x)$ describes the object,  $h(x-t)$ represents the convolution kernel ($PSF$ in LM). The convolution produces an image $I(x)$. The convolution operator equals an integral operator in the 3D image space \cite{sibarita_deconvolution_2005}, denotes as in $\mathbb{R} ^{3}$. In contrast to an ideal convolution, real microscopy image acquisition involves noise and artifacts from measurements, hardware, or inaccurate physical models. They are denoted as the term $\varepsilon$.

\begin{equation}
I(x) = \int_{\mathbb{R}^3} O(x) h(x - t) \, dt + \varepsilon 
\label{convPhysic}
\end{equation}

The convolution kernel $h(x-t)$ relates to the wavelength of the light during imaging. Depending on the optical models, it varies between different microscopy techniques. The Arnison-Sheppard optical model \cite{arnison_3d_2002} modeled the optic convolution as Eq. \ref{Inverse}. In the context of fluorescence microscopy within this study (e.g., widefield microscopy), both the excitation lights wavelength $u_{\lambda ex}$ and emission lights wavelength $u_{\lambda em}$ contribute to the $PSF$ \cite{pawley_handbook_2006}

\begin{equation}
h\left( x,y,z\right) =\left| u_{\lambda ex}\left( x,y,z\right) \right| ^{2}\left| u_{\lambda em}\left( x,y,z\right) \right| ^{2}
\label{Inverse}
\end{equation}

\subsection{Inverse Problem}

Signal measurement models can be reformulated as inverse problems eg. tomography imaging \cite{fan_solving_2019-1}, geodesic measurement \cite{kurylev_rigidity_2010}, and Magnetic Resonance Imaging \cite{song_solving_2022}. Eq. \ref{ResIm} describes the general form of the imaging acquisition system. The $u$ is the true signal. $A$ is defined as the forward operator (e.g. the PSF in this work) and $b$ represents the measured signal. The $\varepsilon$  accounts for the disturbing during measurements. 
\begin{equation}
    b = A(u) + \varepsilon 
    \label{ResIm}
\end{equation}
Solving the inverse problem refers to recovering the signal $u$ from the noisy measurements $b$. Assuming a measurable Hilbert space $\mathbb{R} ^{n}$, it is equipped with a scalar product and normalization. Define two distributions $X$ and $Y$ in this space, the measurements satisfy the relation $b\in Y$ and $u\in X$. Inversion of the forward operator $A$ refers to computing a mapping $Y\rightarrow X$ \cite{heaton_wasserstein-based_2021}. In most cases, the nonlinear property of the forward operator $A$ and the disturbing term $\varepsilon$ introduce the bias in the mapping. It renders the solution of inverse problems neither unique nor stable, which is defined as the ill-posed inverse problem \cite{kabanikhin_definitions_2008, osullivan_statistical_1986}. Instead of directly seeking the mapping $Y \rightarrow X$, a more intuitive approach is to compute the optimization objective $\min |A(u) - b|^2$. To mitigate the solution bias, a practical method involves adding regularization terms $\phi(u)$ to narrow down the search space during computations. This reshapes the optimization process as follows in Eq. \ref{regInverse}. 
\begin{equation}
    \min \left\| A\left( u\right) -b\right\| ^{2}+\phi \left( u\right)
    \label{regInverse}
\end{equation}
The term $\phi(u)$ imposes constraints on the solution space, enhancing the robustness of solutions. Depending on the distinct physical background, regularization methods range from the sparsity regularizer \cite{candes_quantitative_2006, bohning_compressed_2022} and the Tikhonov regularizer \cite{golub_tikhonov_1999} to the total variation regularizer \cite{yin_bregman_2008, bredies_sparsity_2020}. The evolution of NN models also spurred research into the data-driven regularizers \cite{lunz_adversarial_2019}. 

For imaging systems, previous work \cite{fan_solving_2019, feng_levenbergmarquardt_2007, schuster_wavepath_1993} formulated an approximation of the inverse problem below. This approximation contains two operators $K^{T}$ and $\left( K^{T}K+\varepsilon I\right)^{-1}$.
The $K^{T}$ is defined as a forward operator, representing a cluster of math convolutions. The operator $\left( K^{T}K+\varepsilon I\right) ^{-1}$ characterizes a pseudo-differential operator (Eq. \ref{deconvInMatrix1}). The $\overline{u}$ denotes the pseudo restoration target of the true signal $u$.

\begin{equation}
    u \approx \overline{u} = \left( K^{T}K+\varepsilon I\right) ^{-1}K^{T}\overline{b}
    \label{deconvInMatrix1}
\end{equation}

Resolving deconvolution entails calculating the inverse mapping for the operator $A$ between images and the ground truth. The inverse operator is denoted as \(A^{-1}\). Given the ill-posed properties, defining explicit equations for inverse mapping proves challenging. However, Eq. \ref{inverseApproximation} proposed a reasonable approximation for the restoration.

\begin{equation}
    A^{-1} \approx \left( K^{T}K + \varepsilon I \right)^{-1} K^{T}.
    \label{inverseApproximation}
\end{equation}

\subsection{ Beylkin, Coifman, and Rokhlin (BCR) Representation}

Wavelets have been studied for a long time to represent integral and differential operators effectively, e.g. pseudo-differential operators \cite{noauthor_wavelet_nodate} and Calderon-Zygmund operators \cite{meyer_wavelets_1997}. Due to the vanishing moment \cite{cohen_numerical_2003}, direct representation grapples with sparsity issues \cite{daubechies_orthonormal_1988}. This results in computationally intensive operations up to $O\left( N\log N\right)$. The high computational cost diminishes its practical applicability in real-life scenarios. To tackle this issue, Beylkin, Coifman, and Rokhlin (BCR) \cite{beylkin_fast_1991} introduced a cost-effective solution with a nonstandard form of wavelet, reducing the complexity to \(O(N)\).

In the wavelet decomposition theory, a signal can be represented through multi-level decompositions in the space \(v_{1}\). Each decomposition comprises the scaling element and corresponding wavelet coefficient elements (the mother function). For the Daubechies wavelet \cite{daubechies_orthonormal_1988}, the scaling function is denoted by Eq. \ref{Daubechies}.
 \begin{equation}
     \varphi _{k}^{(l)}\left( x\right) =2^{1/2}\varphi \left( 2^{l}x-k\right) ,l=0,1,\ldots 
 \label{Daubechies}
 \end{equation}
The coefficient function is denoted as below (Eq. \ref{coeff_func}).
\begin{equation}
    \psi \left( x\right) =\sqrt{2}\sum _{i \in Z}g_{i}\varphi \left( 2x-i\right) ,g_{i}=\left( -1\right) ^{1-i}h_{1-i}
    \label{coeff_func}
\end{equation}
Given a function $V\left( x\right) $, its scaling and wavelet coefficients can be denoted as Eq. \ref{vk1D} and \ref{psik1D}.
\begin{equation}
    v_{k}^{(l)}:=\int v\left( x\right) \varphi _{k}^{\left( l\right) }\left( x\right) dx,l=0,1,\ldots 
    \label{vk1D}
\end{equation}
\begin{equation}
 d_{k}^{\left( 1\right) }:=\int v\left( x\right) \psi _{k}^{(l)}\left( x\right) dx,l=0,1,\ldots 
    \label{psik1D}
\end{equation}

The graph in Eq. \ref{graph} below illustrates the recursive relation between decomposition levels.

\begin{equation}
\centering 
\includegraphics[width=0.45\textwidth]{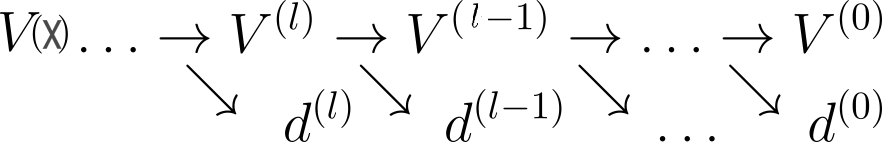} 
\label{graph}
\end{equation}

In practical applications, the process truncates the chain at \(L=L_0\) before the wavelet and scaling functions begin to overlap. This truncation also ensures computationally efficient operations within an affordable range.

Expanding the application to a double integration operator \(A\) for the integration operation \(u=A^{\ast }v\), the recursive relation for double dimension can be denoted as follows in Eq. \ref{doubInt}.\(D_i\) represents complicated integral operators for \(\varphi _{k}^{(l)}\) and \(\psi _{k}^{(l)}\). \(W_i\) are orthogonal operators derived from the identity matrix. For specific definitions and proof, refer to the work \cite{fan_bcr-net_2019}.

\begin{equation}
\includegraphics[width=0.45\textwidth]{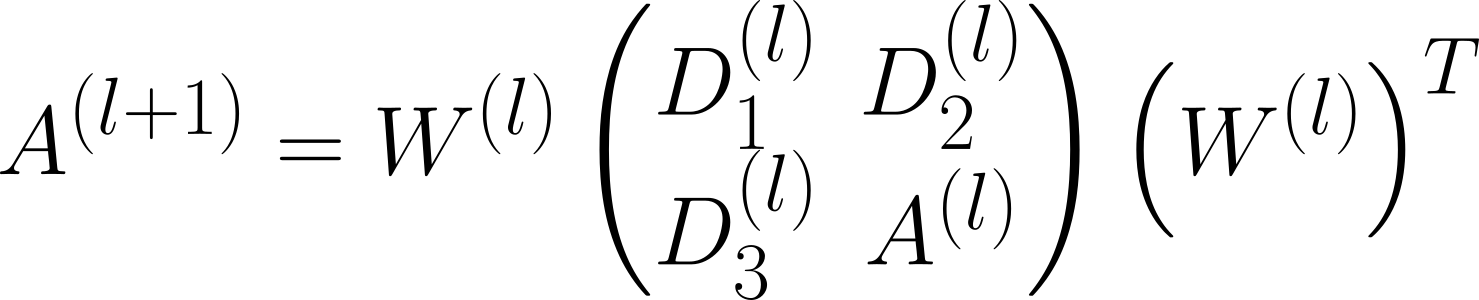} 
\label{doubInt}
\end{equation}

the recursive relation on the integral operator between decomposition level $l+1$ to $l$ can be denoted as in Eq. \ref{recrel}

\begin{equation}
\includegraphics[width=0.53\textwidth]{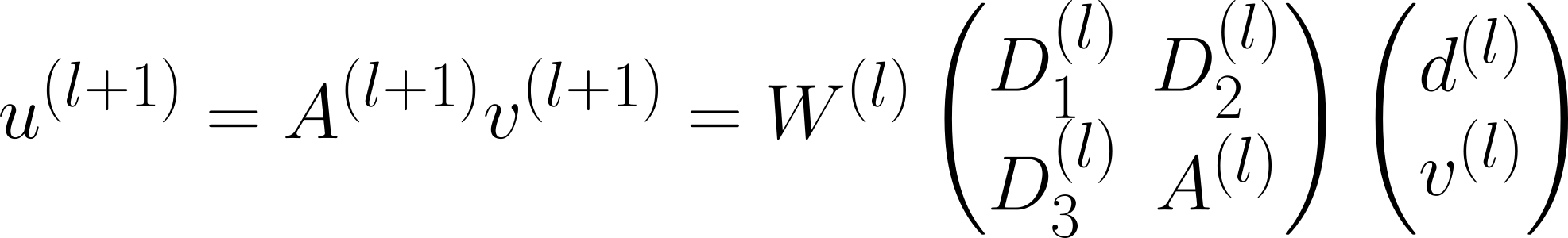}
\label{recrel}
\end{equation}

\begin{figure}[h] 
\centering 
\includegraphics[width=0.65\textwidth]{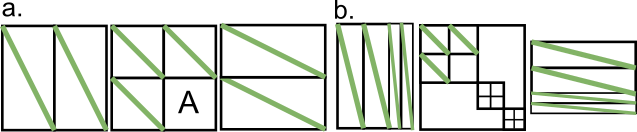}
\caption{BCR-based decomposition. a) presents the decomposed operator \cite{fan_bcr-net_2019} in Eq. \ref{doubInt} at one resolution $l$ with Beylkin, Coifman, and Rokhlin (BCR) wavelet theory. b) illustrates a simple multi-level decomposition based on a). All matrices shown here are band matrices with only values on the green lines.} 
\label{matrixVisual} 
\end{figure}

Till now, we have the decomposition form of the integral operators based on the BCR wavelet theory. To be noticed, all the matrices here are band matrices due to the nonstandard form pre-conditions. For the pseudo-differential operator, we can reshape it to a similar form as in Eq. \ref{pseudoDiff}. It owns a structure akin to the integral operator \cite{ying2022solving, fan_solving_2019}. Therefore, the approximation for pseudo-differential operators can be achieved using the same method as well. For a clearer understanding, Fig.\ref{matrixVisual} visually depicts the matrix operation during the BCR decomposition.

\begin{equation}
    \left( Av\right) \left( x\right) =\int e^{ix\xi }\sigma \left( x,\xi \right) \widehat{v}\left( \xi \right) d\xi =\int a\left( x,y\right) v{(y)},dy
    \label{pseudoDiff}
\end{equation}

\subsection{Neural network approximation}

In a crucial prior study \cite{fan_bcr-net_2019}, the authors connected the BCR decomposition with specific neural network (NN) structures to approximate the linear forward operator $A$. In correspondence with the matrix operation in the recursive chain, the neural network operator $conv[2,2p,2](\xi)$ can replace the multiplication with $\left(W^{(l)}\right)^{T}$. The multiplication with the matrix of $A(l)$ is approximated with local convolution operators $LC[2, nb, 1](\xi)$. In this context, \(nb\) serves as a hyperparameter determining the truncated bandwidth of the band matrices. It represents the entries of local convolution operations in NN. The truncation $u\left( L_0\right) $ can be represented as dense connection $Dense^{2}\left[ 1,1\right] \left( V^{L_0}\right) $. 

At the specific decomposition level \(l\), the original BCR theory can be denoted by the NN structure in Fig. \ref{singleStage}.c without concatenation. This configuration demonstrates robustness in approximating linear systems. Prior studies \cite{fan_solving_2019, fan_solving_2019-1} employed this structure to approximate the tomography physics process. However, the linearity requirements of BCR theory result in the sensitivity of such structure to disturbances during approximation. Given the corrupted microscopy images in this work, the original structure fails to address the strongly non-linear challenges during optic deconvolution.

\section{Methods}
\label{sec:method}

\subsection{Residual BCR Network}

The optical model in Eq. \ref{convPhysic} describes microscopy imaging as a convolution operator in $\mathbb{R} ^{3}$. This can be simplified as $I(t)=A\ast o(t)+\varepsilon$. In Eq. \ref{Inverse}, the deconvolution process is formulated as computing the inverse mapping $I\left( t\right) \rightarrow O\left( t\right)$. Due to ill-posed properties, the inverse operator in Eq. \ref{regInverse} requires regularizers to ensure robustness and address challenges related to uniqueness.

As explained in Eq. \ref{pseudoDiff}, the inverse operator of an integral can be formulated as a pseudo-differential operator ($A^{-1}$). This can be represented as an approximation using BCR theory. At resolution level $l$, the convolution is decomposed as $I^{(l)}=A^{(l)}o^{(l)}+\epsilon^{(l)}$. For simplicity, we assume the noise term $\epsilon^{(l)}$ as a constant $\varepsilon_0$ at all levels. Thus, we can describe the deconvolution as Eq. \ref{deconvMatrix}. 
\begin{equation}
    o^{\left( l\right) }=\left( A^{(l)}\right) ^{-1}\left( I^{\left( l\right) }-\varepsilon ^{\left( l\right) }\right) \\
    =\left( A^{(l)}\right) ^{-1}I^{\left( l\right) }+\varepsilon _{0}
    \label{deconvMatrix}
\end{equation}
Introduce the recursive relation in Eq. \ref{recrel} into Eq. \ref{deconvMatrix}, the integral operator then can be denoted as Eq. \ref{Diagram}

\begin{equation} 
\centering 
\includegraphics[width=0.5\textwidth]{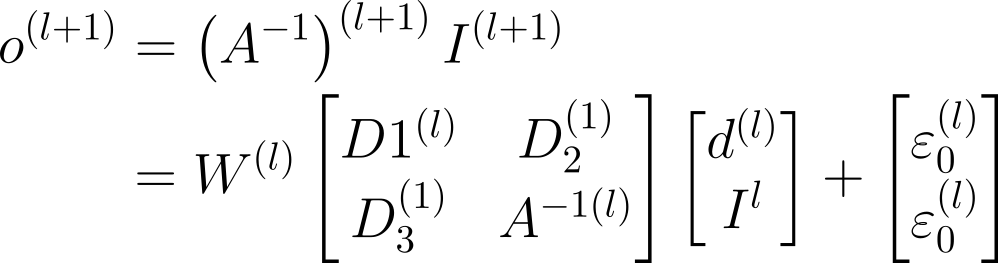}
\label{Diagram} 
\end{equation}

We adopted the Eq. \ref{deconvInMatrix1} as the backbone of our models. The deconvolution process is then described as Eq. \ref{deconvInMatrix2}. To be noticed, the first $I$ indicates the identity matrix from Eq. \ref{recrel}. The second $i$ is a microscopy image in this work in Eq. \ref{deconvMatrix}.
\begin{equation}
    o \approx \overline{o}= \left( K^{T}K+\varepsilon I\right) ^{-1}K^{T}i
    \label{deconvInMatrix2}
\end{equation}

\begin{figure}[h!] 
\centering 
\includegraphics[width=0.85\textwidth]{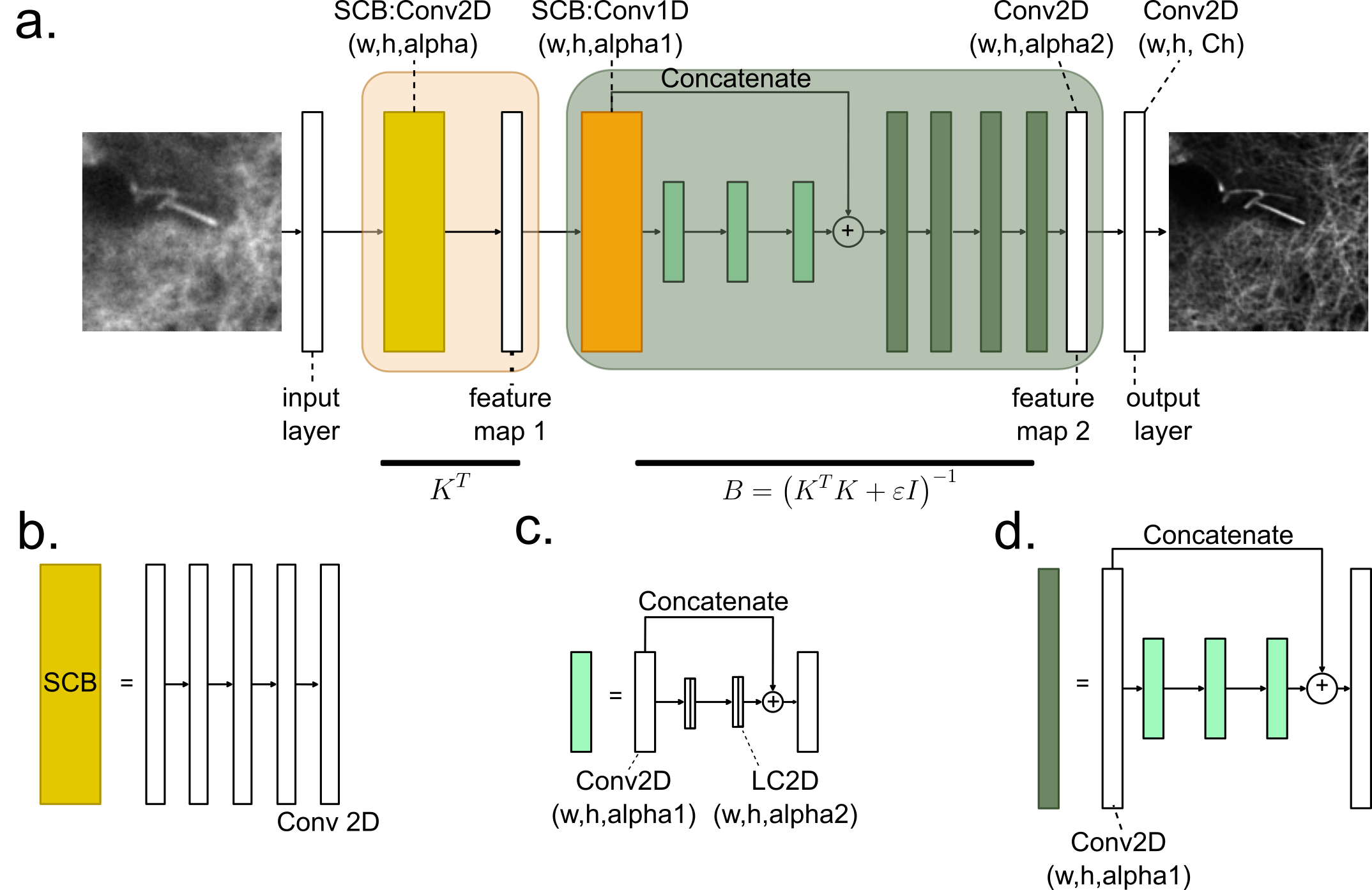}
\caption{Single stage residual BCR Net. a) indicates the structure of the NN model for microscopy deconvolution. it consists of the Sequence Convolution Block (SCB) in b) as $K^{T}$ and pseudo-differential operator from sub-units of c) and d). c) illustrates the regularised BCR decomposition through the residual structure. } 
\label{singleStage} 
\end{figure}

The $K^{T}$ represents a cluster of convolutions in a neural network (NN), and $B = (K^{T}K + \varepsilon I)^{-1}$ approximates a pseudo-differential operator. Nevertheless, the model exhibits sensitivity to noise. Since noise and artifacts are common in microscopy image acquisition, the original NN structure is extremely unstable for microscopy deconvolution. As indicated by Eq. \ref{regInverse}, this model requires additional regularization to stabilize the learning process. The continuous residual NN structure has shown strength in providing robustness in optimization \cite{chen_neural_2018} in previous computer vision work \cite{zhang_residual_2018}. To enhance the model's robustness by nonlinear problems, we proposed the BCR sub-module with the residual dense structure \cite{zhang_residual_2018} in Fig. \ref{singleStage}.c. It concatenates the images to the output and introduces constraints to regularize the learning. The recursive relation is then reshaped as $o^{(l+1)} = (A^{-1})^{(l+1)} (I^{(l+1)} | I^{(l)})$. Based on this, we propose a vanilla demo Single-Residual BCR Net (s-rBCR) in Fig. \ref{singleStage}.a. Till now, both operators in Eq. \ref{recrel} in a strongly nonlinear microscopy system can be approximated from the structures in Fig. \ref{singleStage}.b and Fig. \ref{singleStage}.c.

\begin{figure}
\centering 
\includegraphics[width=0.85\textwidth]{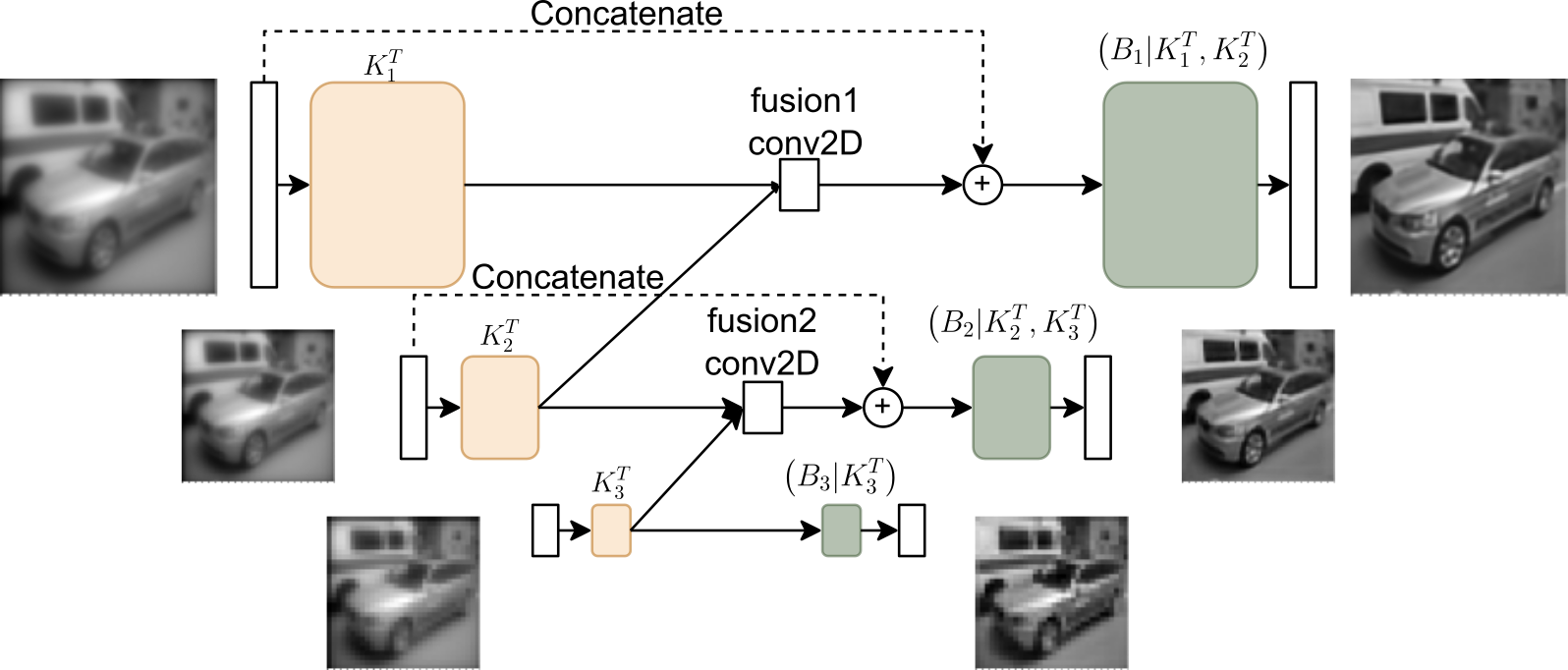}
\caption{Multi-Stage Residual BCR Net (m-rBCR). The m-rBCR model reintegrates the input image from other resolutions as posterior for the pseudo-differential operator $B = \left( K^{T}K + \varepsilon I \right) ^{-1}$.}  
\label{multiStage} 
\end{figure}

The original BCR theory truncated decomposition levels to $L_0$ for computational efficiency, resulting in an accumulated loss during deconvolution. To address this, we reintegrate the input as posteriors at each resolution level using Multi-stage learning \cite{fichtinger_multi-stage_2011} \cite{zeng_multi-stage_2013}. Fig.\ref{multiStage} illustrates our proposed multi-stage residual BCR net (m-rBCR). In m-rBCR, Pseudo-differential operators with posteriors are denoted as $\left(  B_{1}| K_{1}^{T},K_{2}^{T}\right) $ and $\left(  B_{2}| K_{2}^{T},K_{3}^{T}\right) $.

The pseudo-code in Algorithm \ref{alg:forward} and \ref{alg:pseudoDiff} illustrates the NN structure for the single-stage residual BCR Net. Based on Eq. \ref{deconvInMatrix2}, the deconvolution consists of two parts: the forward operator $K^{T}$ and the pseudo-differential operator $B =\left( K^{T}K+\varepsilon I\right) ^{-1}$. 

For the hyperparameters of $K^{T}$ in Algorithm \ref{alg:forward}, $n_b$ represents the bandwidth of the truncated matrix \cite{fan_bcr-net_2019}, also the entries of the local convolution in NN. $L$ and $L_0$ stand for the decomposition level and truncation level in Eq. \ref{graph}. $RDN$ represents the depth number of the residual dense block (RDN). We recast the input as a posterior to regularize the learning at code line $15$. Algorithm \ref{alg:pseudoDiff} illustrates the pseudo-differential operator $B =\left( K^{T}K+\varepsilon I\right) ^{-1}$. $\alpha_i$ and $w_i$ stand for the configuration of the NN structure. $L_i$ represents the recovery level in the previous theory. $RDN$ represents the depth. The residual block complies with the structure proposed in Fig. \ref{singleStage}.b. The m-rBCR model derives from the base structure s-rBCR. As in Fig.\ref{multiStage}, We added the feature fusion from other resolutions and current inputs to recover the truncated signal.

\begin{figure}[tb]
\SetAlFnt{\small}
\SetAlCapFnt{\small}
\SetAlCapNameFnt{\small}

\begin{minipage}[t]{0.52\linewidth}
\begin{algorithm}[H]
\caption{The forward operator $K^{T}$}
\label{alg:forward}
\SetKwInOut{Input}{Input}
\SetKwInOut{Output}{Output}

\Input{$input, \alpha, n_b, K, L, L_0, RDN$}
\Output{Feature map  $u_{map}$}

x-in = $input$, depth = $RDN$ \;
v = [None], v[L] = x-in \;
x = [None] \;

\For{$l$ from $L-1$ to $L_0-1$ by -1}{
    $x[l] = \text{Residual-Dense-Block}(v[l+1], \text{depth}, n_b)$\;
    $v[l] = x[l][..., \alpha]$}
    
u = [None] * ($K$ + 1) \;
u[0] = v[$L_0$] \;

\For{$k$ from 1 to $(K+1)$ by 1}{
    u[k] = $Dense-2D$(u[$k-1$])}

\For{$l$ from $L_0$ to $L$}{
    $y = x[l]$ \;
    $y = \text{Residual-Dense-Block}(v[l+1], n_b)v[l]$ \;
    \For{$k$ from 0 to $K$}{
        $z = \text{LocallyConnected-2D}(y)\text{ \;}$}
    \tcp{recap posterior}
    Fusion = $\text{Concatenate}(z, u[l])$ \;}

$u_{map}$ = $Reshape$(Fusion)

\end{algorithm}
\end{minipage}
\hfill
\begin{minipage}[t]{0.48\linewidth}
\begin{algorithm}[H]
\caption{The pseudo-differential operator $B =\left( K^{T}K+\varepsilon I\right) ^{-1}$}
\label{alg:pseudoDiff}
\SetKwInOut{Input}{Input}
\SetKwInOut{Output}{Output}

\Input{$Feature, \alpha_1, \alpha_2, w_1, w_2,$\\ $L_1, L_2, RDN$}
\Output{deconvolution $I_{deconv}$}

x-in = $Feature map$ \; 
depth = $RDN$ \;
x = x-in \;
\For{$l$ in $L_1$}{
    x = $Conv-1D(\alpha_1, w_1)(x)$\;} 
    
\For{$l$ in ($L_2$)}{
    x = Residual-Dense-Block($\alpha_2, w_2, depth$)(x)\;}
$I_{deconv}$ = $Reshape$(x) \;
\end{algorithm}
\end{minipage}
\end{figure}

\subsection{Simulated Dataset for Benchmarking}
Based on the optical model above, we adopted the widefield microscopy simulation model (BioSR and ImageNet as the ground truth) from other works \cite{li_microscopy_2023}. The model contains PSFs with physically plausible ranges for microscopy hardware parameters: the numerical aperture ranged between 0.4 and 1.0, the excitation wavelength between 450 nm and 490 nm, and the emission wavelength between 450 nm and 600 nm. The Pinhole size was set to 0.1 \(\mu\)m for the confocal microscopy and 1000 \(\mu\)m to simulate the absent (open) pinhole situation for the widefield microscopy in the dataset. The refractive index was fixed at 1.33, corresponding to the water refractive index. These diverse PSFs were then utilized for convolution. We took two datasets: a biological dataset BioSR \cite{noauthor_evaluation_nodate} and ImageNet \cite{deng_imagenet_2009}. We augmented the generated images by adding Gaussian noise (with $\sigma$ ranging from 0.01 to 0.05) and Poisson noise (with $\lambda$ ranging from 0.01 to 0.05) to enhance comparability to real microscopy images. All image values were rescaled to the [0, 1] range. The dataset consists of over 57k images. We split it into training/validation/test with the widely used ratio 0.8/0.1/0.1. 


\section{Experiments}
\label{sec:experiment}

In our experiments, we compared 10 optical deconvolution models (Tab.\ref{Evaluate}). The Richardson-Lucy (RL) model relies on the explicit definition of the PSF, while the rest are NN-based. For the NN-based models, this work covers most of the state-of-the-art models of denoising/deblurring. Specifically, the U-Net \cite{navab_u-net_2015}, CARE \cite{weigert_content-aware_2018}, DnCNN \cite{zhang_beyond_2017}, Noise2Noise \cite{lehtinen_noise2noise_2018}, MIMO-U-Net \cite{cho_rethinking_2021}, RCAN \cite{zhang_image_2018}, MPRNet \cite{zamir_multi-stage_2021}, ESRGAN \cite{wang_esrgan_2018} and DDPM \cite{ho_denoising_2020} with U-Net as denoising substructures. Notably, our s-rBCR and m-rBCR are specifically designed for inverse convolution in microscopy optical models. For each experiment, we report the number of parameters, runtime in seconds, peak signal-to-noise ratio (PSNR), and structural similarity index (SSIM).

\begin{table}[ht]
  \caption{Models evaluation. Test on two simulated widefield datasets (BioSR and ImageNet) and two real 
 widefield datasets (widefield-confocal(W-C) and dSTORM). Parameters in units million and runtime in seconds. Sim. indicates simulated dataset.}
  \label{table_transposed}
   \begin{adjustbox}{width=.95\columnwidth,center}
    \begin{tabular}{llllllllll}
      \toprule
      Model & Params. ($\uparrow$) & Runtime & \multicolumn{2}{c}{BioSR(sim.)} & \multicolumn{2}{c}{ImageNet(sim.)} & \multicolumn{2}{c}{W-C(real)}& \multicolumn{1}{c}{dSTORM(real)} \\
      \cmidrule(lr){4-5} \cmidrule(lr){6-7} \cmidrule(lr){8-9} \cmidrule(lr){10-10}
      & & & PSNR & SSIM & PSNR & SSIM & PSNR & SSIM & \multicolumn{1}{c}{PSNR} \\
      \midrule
      RL & N/A & 0.0835 & 13.39 & 0.48 & 12.74 & 0.78 & N/A & N/A & \multicolumn{1}{c}{N/A} \\
      s-rBCR & 0.086 (ours) & \textbf{0.0019} & 20.28 & 0.59 & 20.84 & 0.83 & 17.67 & 0.57 & \multicolumn{1}{c}{19.55} \\
      m-rBCR & 0.237 (ours)  & \underline{0.0036} & \textbf{24.89} & \textbf{0.78} & \underline{21.41} & \underline{0.86} & \textbf{23.10} & \underline{0.70} & \multicolumn{1}{c}{\textbf{20.13}} \\
      CARE & 0.333 & 0.0039 & 22.15 & 0.65 & 18.47 & 0.74 & 21.53 & 0.66 & \multicolumn{1}{c}{19.62} \\
      DnCNN & 0.556 & 0.0056 & 21.41 & 0.70 & 19.70 & 0.84 & 19.34 & 0.63 & \multicolumn{1}{c}{17.46} \\
      Noise2Noise & 1.227 & 0.0068 & 16.07 & 0.57 & 16.24 & 0.60 & 15.07 & 0.57 & \multicolumn{1}{c}{18.06} \\
      MIMO-U-Net & 6.807 & 0.0087 & \underline{23.95} & \textbf{0.78} & \textbf{22.35} & \textbf{0.88} & 19.17 & 0.67 & \multicolumn{1}{c}{18.91} \\
      U-Net & 7.780 & 0.0241 & 21.89 & \underline{0.73} & 19.23 & 0.75 & 18.17 & 0.63 & \multicolumn{1}{c}{19.62} \\
      RCAN & 15.334 & 0.0281 & 21.71 & 0.64 & 19.78 & 0.91 & 20.51 & 0.58 & \multicolumn{1}{c}{19.26} \\
      MPRNet & 20.127 & 0.0173 & 21.44 & 0.63 & 20.12 & 0.83 & 21.53 & 0.55 & \multicolumn{1}{c}{18.54} \\
      DDPM $L_2$ & 23.988 & 1.0968 & 21.85 & 0.65 & 20.22 & 0.78 & \underline{22.27} & \textbf{0.73} & \multicolumn{1}{c}{17.46} \\
      ESRGAN & 49.841 & 0.0147 & 19.82 & 0.59 & 18.95 & 0.76 & 21.17 & 0.59 & \multicolumn{1}{c}{\underline{19.93}} \\
      \bottomrule
    \end{tabular}
    \end{adjustbox}
  \label{Evaluate}
\end{table}

As depicted in Tab.\ref{Evaluate}, the BCR family models are notably slimmer compared to the other models. ESRGAN, the largest model, surpasses our model m-rBCR net by a factor of 200 times. The MIMO-U-Net, the current benchmark in image deblurring, is approximately 30 times larger than m-rBCR. The RL model is deterministic and does not contain trainable parameters. An inherent advantage of lightweight models is the efficiency of runtime testing. We evaluated all the models on the BioSR test dataset of 500 images. The s-rBCR demonstrated the highest speed, while the m-rBCR exhibited the second-fastest performance. The m-rBCR surpassed the benchmark MIMO-U-Net by 3 times speed improvement.

All models were trained on simulated datasets with explicit PSF information to master optical deconvolution. Subsequently, we tested them across four datasets: simulated BioSR, simulated ImageNet, real widefield-dSTORM microscopy image pairs \cite{noauthor_evaluation_nodate}, and real widefield-confocal image pairs \cite{pawley_handbook_2006}). In the two real microscopy datasets, we utilized the dSTORM images and confocal microscopy images as the ground truth. For evaluations, we computed the PSNR and SSIM on the test results.

\subsection{Benchmark on a Simulated Dataset}

\begin{figure*}
  \centering 
    \includegraphics[width=0.9\textwidth]{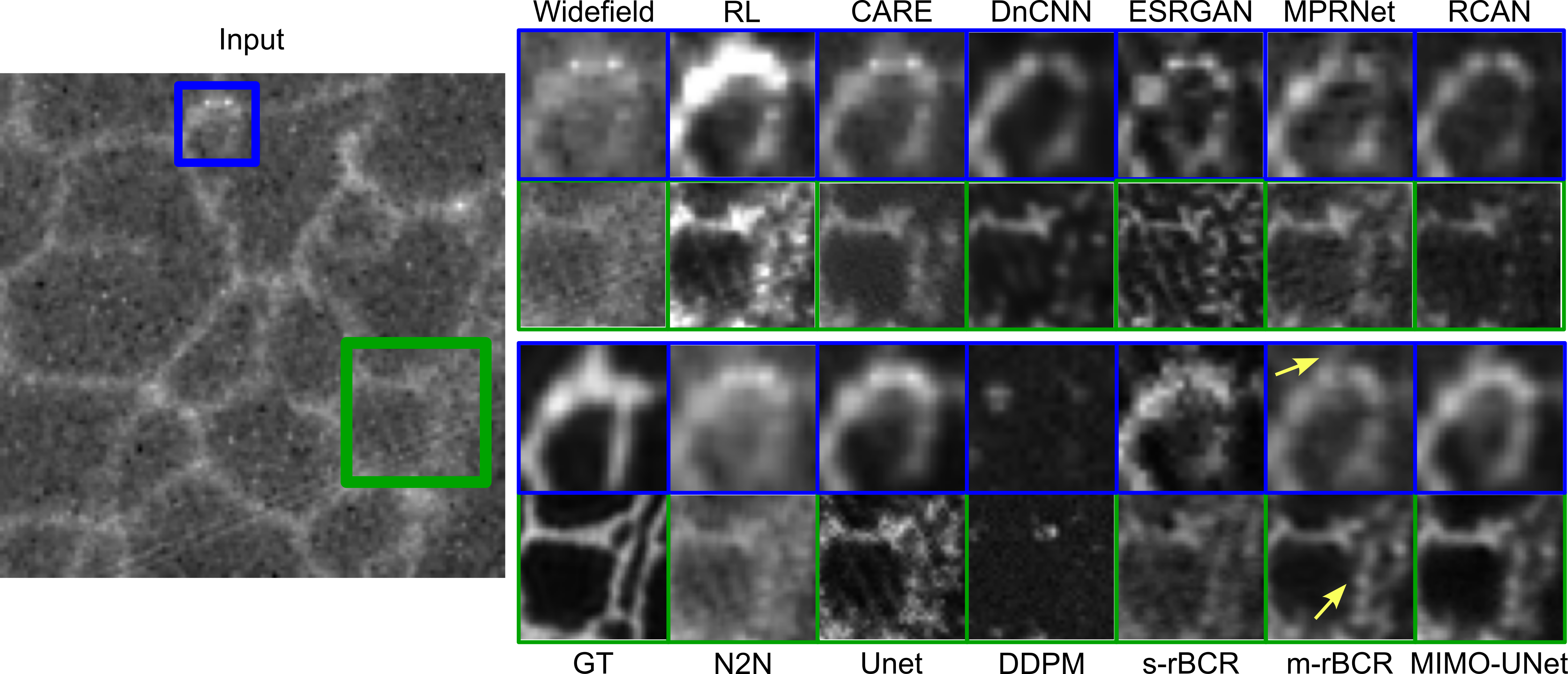}
    \rule{.9\linewidth}{0pt}
    \caption{Test results on the simulated widefield microscope dataset from BioSR. Our m-rBCR model successfully restored comparable details to the benchmark MIMO-U-Net, demonstrating efficiency with fewer parameters and a shorter runtime.}  
    \label{bioResult} 
  \hfill
\end{figure*}

Fig. \ref{bioResult} presents part of the results from the deconvolution by the BioSR test set. Despite the strong blur observable in the widefield microscopy input images, our m-rBCR model effectively restored detailed information. Notably, the m-rBCR successfully recovered high-frequency information without introducing additional artifacts in the region indicated by the arrow (Fig. \ref{bioResult}). Table \ref{Evaluate} provides a comprehensive evaluation of the models' performance on the BioSR test set. Despite utilizing an explicit PSF, the RL model exhibited the lowest rankings (PSNR: 13.39, SSIM: 0.48) attributable to noise disturbance. Among blind-deconvolution models, our m-rBCR outperformed (PSNR: 24.89, SSIM: 0.78) the others, while MIMO-U-Net ranked on second place (PSNR: 23.95, SSIM: 0.78). To MIMO-U-Net, m-rBCR utilizes 30 times fewer parameters and completes the runs 3 times quicker, serving as the performance upper bound.

\begin{figure*}
  \centering 
    \includegraphics[width=0.95\textwidth]{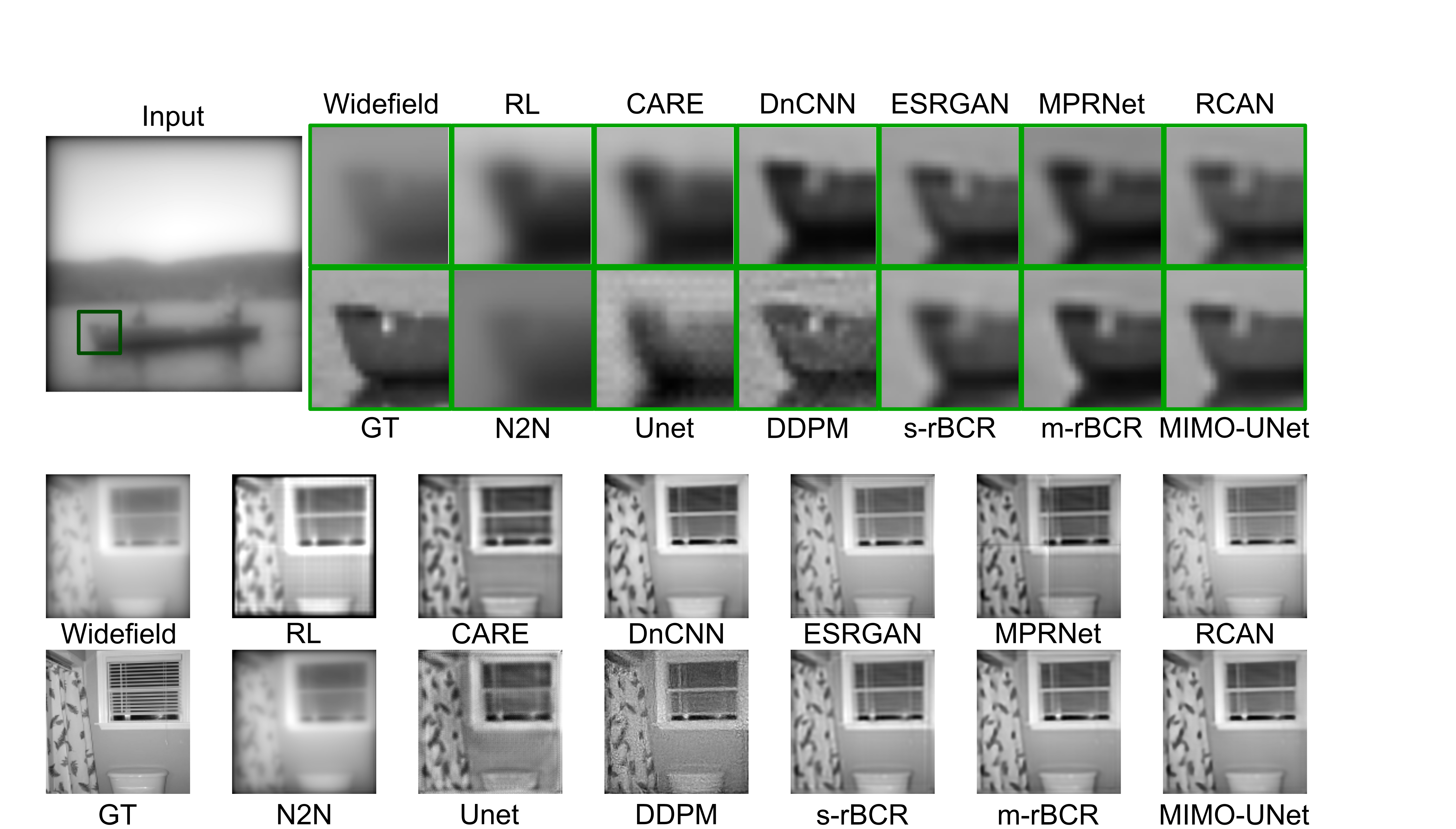}
    \rule{.9\linewidth}{0pt}
    \caption{Test results on the simulated widefield microscope dataset from Imagenet. Multi-stage residual BCR Net (m-rBCR) restores the most details without introducing additional artifacts due to noise.}  
    \label{imnResult}
  \hfill
\end{figure*}

Fig. \ref{imnResult} illustrates the results on the ImageNet test set. While the DDPM model provides the sharpest restoration, it introduces a strong noisy background texture. In contrast, the deconvolution by the MIMO-U-Net and m-rBCR models exhibits significantly fewer artifacts while successfully restoring rich details. The complete evaluation is outlined in Table \ref{Evaluate}. The m-rBCR model attains the second-best performance (PSNR: 21.41, SSIM: 0.86), following the benchmark MIMO-U-Net (PSNR: 22.35, SSIM: 0.88).

\subsection{Deconvolution of Experimental dSTORM Microscopy Data}

\begin{figure*}
  \centering 
  \includegraphics[width=0.95\textwidth]{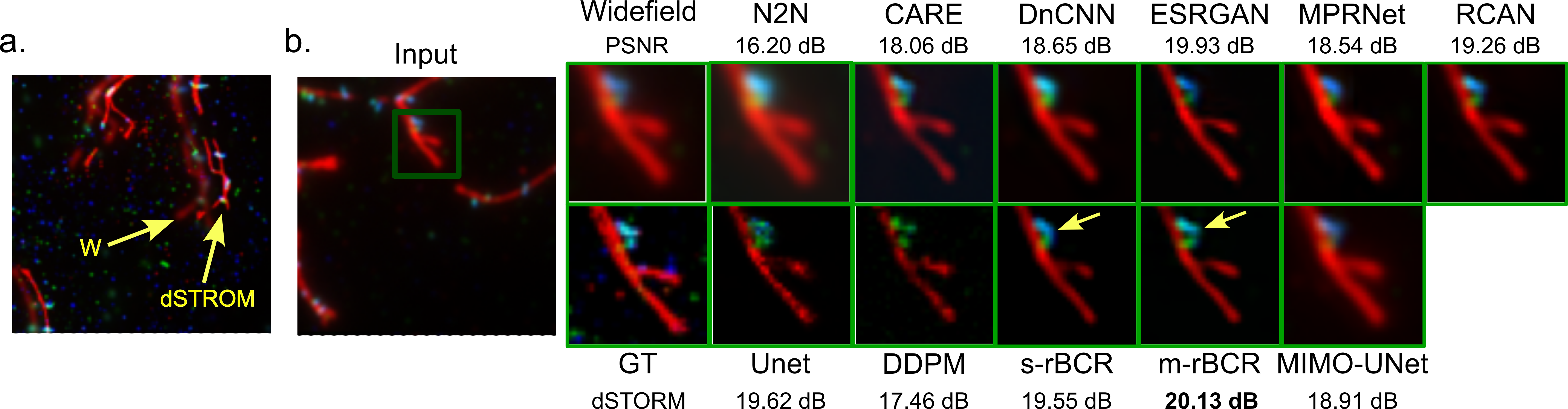}
    \rule{.9\linewidth}{0pt}
  \caption{Test results on the experimental dataset from dSTORM.
    a) The drift and distortion during the measurements (widefield-dSTORM). Thus, the SSIM is invalid to measure the deconvolution quality.
    b) Presents part of the results from different deconvolution models. The m-rBCR (PSNR-20.13 dB) ranks at the top.
  }
  \label{stormResult.}
  \hfill
\end{figure*}

We validated our models on real dSTORM microscopy images in Fig.\ref{stormResult.} This publically available dataset contains widefield microscopy images and the corresponding super-resolution Direct Stochastic Optical Reconstruction Microscopy (dSTORM) dataset \cite{noauthor_three-color_nodate}. Due to the lack of ground truth in real microscopy, the dSTORM serves as the pseudo-ground truth. Since the PSF is not available, we could only perform blind deconvolution without the RL method. As illustrated in Fig.\ref{stormResult.}.a, during real microscopy data collection from two different microscopy techniques, the sample drifted and distorted. Yet, the application of SSIM requires the input signals to be properly aligned \cite{wang_image_2004}. Thus, we did not evaluate SSIM to avoid bias by analysis. The assessment results are presented in Table \ref{Evaluate}.

The performance of MIMO-U-Net declined (PSNR: 18.91 dB), whereas the m-rBCR outperformed the rest in this test set (m-rBCR: 20.13 dB). Notably, both m-rBCR and s-rBCR uniquely restored pixels from extremely weak inputs (arrow indicates). A possible reason is that while expertise-oriented NN models demonstrate solid performance in resolution enhancement, they neglect the physics texture of the process. Without the physical constraints, the models learned features irrelevant to the deconvolution process. This makes the restoration of these models by real microscopy unstable and tends to generate artifacts from those features.

\subsection{Deconvolution of Widefield/Confocal Microscopy Data}

\begin{figure*}[htb]
  \centering 
    \includegraphics[width=0.95\textwidth]{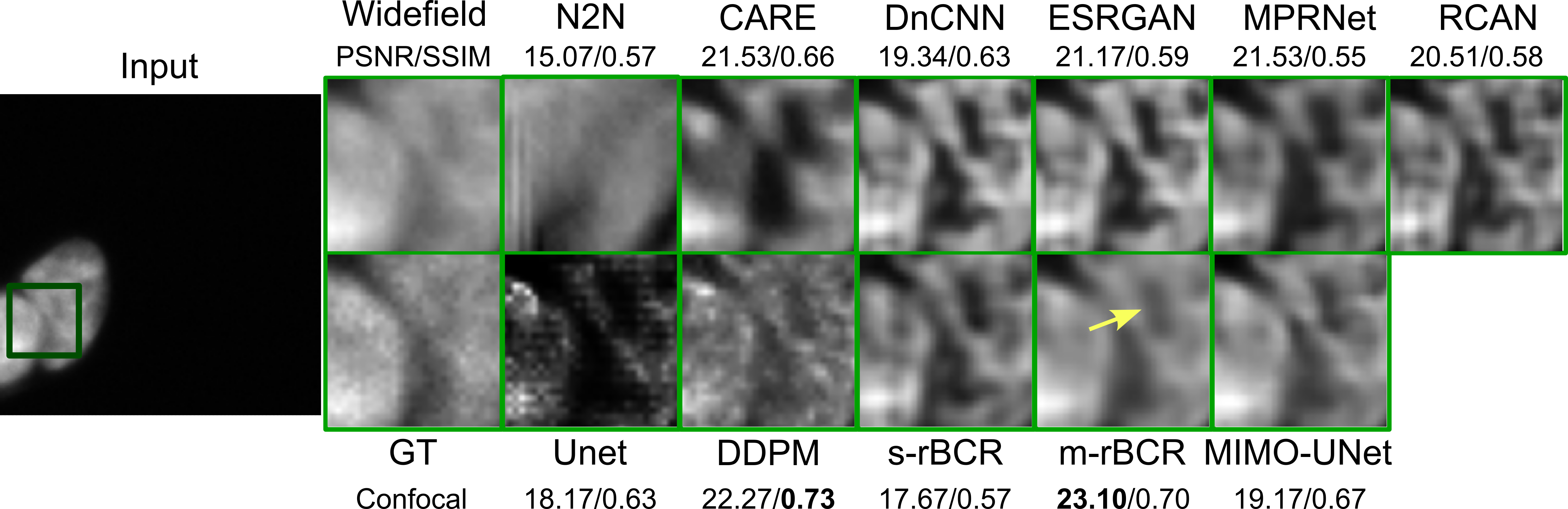}
    \rule{.9\linewidth}{0pt}
    \caption{The deconvolution on real widefield microscopy data. Confocal images serve as pseudo-ground truth. The restorations of m-rBCR (PSNR: 23.10 dB) unveiled previously unrecognizable patterns in the widefield inputs). }  
    \label{w_cResult.}
  \hfill
\end{figure*}

Finally, we evaluated our model on another real microscopy dataset \cite{li_microscopy_2023} consisting of widefield microscopy images with corresponding confocal images, which were used as pseudo-ground truth. The detailed evaluation is presented in Table \ref{Evaluate}. As shown in Fig. \ref{w_cResult.}, the restoration brought out previously unrecognizable details in the input pixels. The m-rBCR took the lead with the highest PSNR of 23.10 dB, followed by the DDPM with a PSNR of 22.27 dB. Notably in the arrow-indicated area, m-rBCR recovers image details without introducing artifacts as in DDPM.

\subsection{Robustness analysis on hyper-params of m-rBCR}

\begin{figure}[htb]
  \centering
  \includegraphics[width=0.95\textwidth]{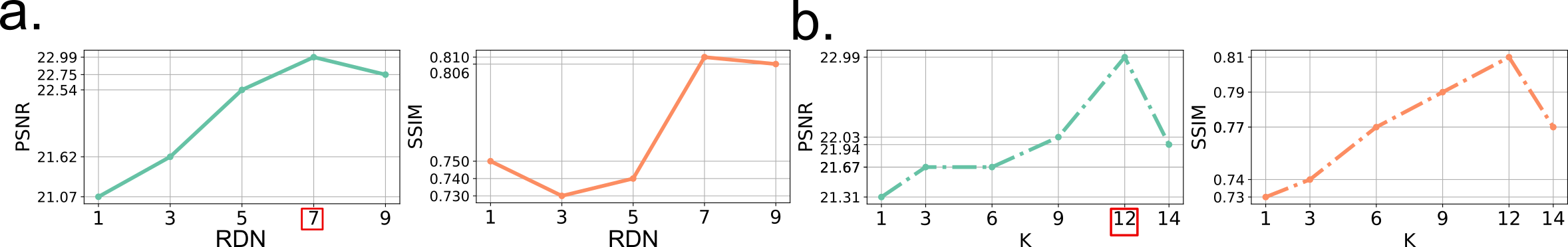}
  \caption{Robustness test on decomposition level $K$ and residual block depth $RDN$ in algorithm \ref{alg:forward} and \ref{alg:pseudoDiff}. Performance peaks at $K=12$, after which it starts to decline. Based on these results, we adopted $RDN=7$ and $K=12$ for this work.
  }
  \label{stableTest}
\end{figure}

Built upon the foundation of explicit physical modeling, the physics-informed m-rBCR model offers enhanced transparency for tuning. In the prior theory, \( K \) signifies the wavelet approximation's decomposition level, while truncation in BCR theory happens at \( K=L_0 \). Overestimating \( K \) leads to excessive decomposition, causing wavelets and scaling functions to overlap. This results in increased computational costs and performance degradation. In the residual structure, the depth of the residual block \( RDN \) influences learning constraints. Intense constraints can greatly bulk up the model, hampering its overall learning ability without performance gains. Thus, we performed numerical robustness experiments regarding $K$ and $RDN$ in the algorithms \ref{alg:forward} and \ref{alg:pseudoDiff} on the simulated BioSR dataset. As shown in Fig.\ref{stableTest}, the performance peaks at the configuration $RDN=7$ and $K=12$. We adopt these parameters in this paper. It is important to note that the configuration may vary depending on the disturbance $\varepsilon_0$ in Eq. \ref{deconvMatrix}.

\section{Conclusions \& Discussion}
\label{sec:conclusion}

In this study, we formulated optic deconvolution in light microscopy as an inverse problem and addressed it by solving a pseudo-differential operator. Inspired by the BCR wavelet approximation theory, we introduce the multi-stage residual BCR net (m-rBCR) to approximate the inverse operation. The m-rBCR adapts to the strong non-linear properties in noisy restoration tasks of microscopy. We validated our m-rBCR model on four test sets: two simulated widefield microscopy datasets (Imagenet and BioSR) and two real microscopy datasets (dSTORM and confocal datasets). To evaluate the performance, we compared the PSNR and SSIM of the m-rBCR's deconvolution results with those from the classical Richardson-Lucy deconvolution model and the state-of-the-art NN-based models by denoising/deblurring (DDPM, U-Net, CARE, DnCNN, ESRGAN, RCAN, Noise2Noise, and MPRNet). Our m-rBCR model secured the top position in performance across the simulated BioSR test set and two real microscopy test sets. In the simulated ImageNet microscopy test set, m-rBCR achieved the second-highest ranking, surpassing other candidates and closely following the benchmark MIMO-U-Net. However, by leveraging the inverse problem framework and the optical model, the m-rBCR model requires 30 times fewer parameters compared to MIMO-U-Net while achieving comparable performance. The direct benefits include m-rBCR demanding significantly fewer computational resources and exhibiting a greatly shorter runtime(from $\sim$3 times faster than MIMO-U-Net to $\sim$300 times than DDPM). This work showcases the benefits of the use of physics models to exploit redundant parameters of NN models and reach powerful performance by microscopy optic deconvolution.

\section{Limitations}
Currently, m-rBCR involves a straightforward pixel mixture from other resolution levels. In the follow-up to our work, the fusion strategies based on wavelet theory could be explored. This could further improve the learning performance.

\section{Code Availability}
Please refer to this repository for the source code:\\ \href{https://github.com/leeroyhannover/m-rBCR}{https://github.com/leeroyhannover/m-rBCR}

\section*{Acknowledgements}
\label{sec:ack}

This work was partially funded by the Center for Advanced Systems Understanding (CASUS) which is financed by Germany’s Federal Ministry of Education and Research (BMBF) and by the Saxon Ministry for Science, Culture, and Tourism (SMWK) with tax funds on the basis of the budget approved by the Saxon State Parliament. MK was supported by the Heisenberg award from the DFG (KU 3222/2-1), as well as funding from the Helmholtz Association. The authors thank HelmholtzAI (grant tomoCAT).

\bibliographystyle{splncs04}
\bibliography{egbib}
\end{document}